\documentclass[3p,times,twocolumn]{elsarticle}

\usepackage{ecrc}

\volume{00}

\firstpage{1}

\journalname{Nuclear and Particle Physics Proceedings}

\runauth{}

\jid{nppp}

\jnltitlelogo{Nuclear and Particle Physics Proceedings}

\usepackage{amssymb}

\usepackage[figuresright]{rotating}

\begin{document}

\begin{frontmatter}



\dochead{}

\title{Cosmic Ray (Stochastic) Acceleration from a Background Plasma}


 \author[label1,label2,label5]{V. A. Dogiel\footnote{Tel.: +7 499 132 6235; Fax: +7 499 135 8533; E-mail address: dogiel@td.lpi.ru} }
\address[label1]{P.N.Lebedev Physical Institute, Moscow, Russia}
\author[label2]{ K. S. Cheng}
\address[label2]{Department of Physics,University of Hong Kong, Hong Kong, China}
 \author[label1,label2]{ D. O. Chernyshov}
 \author[label1,label3]{A. D. Erlykin}
\address[label3]{Department of Physics, Durham University, Durham, UK}
\author[label4]{C.-M. Ko} 
\address[label4]{National Central University,
Zhongli Dist., Taoyuan City, Taiwan (R.O.C.)}
 \author[label3]{and A. W. Wolfendale}
\address[label5]{Moscow Institute of Physics and Technology (State University),  Dolgoprudny, 141707, Russia}



\begin{abstract}
We give a short review of processes of stochastic acceleration in the Galaxy. We discuss: how to estimate correctly the number of accelerated particles, and at which condition the stochastic mechanism is able to generate power-law nonthermal spectra. We present an analysis of stochastic acceleration in the Galactic halo and discuss whether this mechanism can be responsible for production of high energy electrons there, which emit gamma-ray and microwave emission from the giant Fermi bubbles. Lastly, we discuss whether the effects of stochastic acceleration can explain the CR distribution in the Galactic disk (CR gradient).
\end{abstract}


\end{frontmatter}


\section{Introduction}
\label{}
 The theory of CR origin started from the  key papers of Baade and  Zwicky \cite{baade} and Fermi \cite{fermi1,fermi2}. In the first paper the authors assumed that the bulk of CRs observed near Earth are produced by supernovae explosions in the Metagalaxy. They excluded that supernovae in our Galaxy were sources of CRs because, according to their estimates, their  energy density had to be too high. An alternative explanation was suggested by Fermi. He assumed that CRs were of the Galactic origin and the whole volume of the Galaxy was a source of CRs. CR acceleration is in this case due to regular collisions of charged particles with chaotically moving magnetic fluctuations. The acceleration is due to the induced electric field $\mathcal{E}$
\begin{equation}
curl~ \mathcal{E}=-\frac{1}{c}\frac{d{\bf H}}{dt}
\end{equation}
excited by a time-varying magnetic field  $\bf{H}$.

As a result  the energy of particles, $E$, increases as 
\begin{equation}
\frac{dE}{dt} \sim\left(\frac{u}{c}\right)^2\frac{E}{L/c}=\alpha_0E
\end{equation}
where $u$ is the velocity of magnetic fluctuations and $L$ is the average distance between fluctuations.

An advantage of this model was that this mechanism generates power-law spectra of accelerated particles, just as needed for the observed CR power-law spectrum. However, spectra in the Fermi model were much harder ($\propto E^{-1}$ in the limit case) than the observed for CR spectrum. Besides,  too a very time was necessary to accelerate particles up to the  high energies needed.

In 1964 Ginzburg and Syrovatskii \cite{ginzburg} suggested their theory of CR origin. Using the observed CR chemical composition they estimated the CR luminosity, which was about $10^{40}$ erg s$^{-1}$, and concluded  that probably Galactic supernovae are the sources of CRs, because only 10\% of supernovae shocks energy was needed to generate the Galactic CR flux. They developed also the diffusion model of CR propagation in the Galaxy where particle scattering by magnetic fluctuations in the interstellar medium  was described as spatial diffusion.

The next important mile-stone was connected with the papers of Krymskii and Bell \cite{krymskii,bell}  who suggested the theory of CR acceleration by supernovae shocks. In principle, this model is a modification of the classical model of Fermi, because the important component of the acceleration process is that of particles scattering on moving magnetic fluctuations but the new component of the model is a velocity "jump" on the shock front that changes the acceleration process drastically. First, this acceleration generates steeper spectra of particles ($\propto E^{-2}$) just as needed to explain the observed CR spectrum. Secondly the rate of acceleration is proportional to the first degree of $u/c$
\begin{equation}
\frac{dE}{dt}\propto \frac{u}{c}E
\end{equation}
that makes the shock acceleration much more effective than that of Fermi. Here $u$ is the shock velocity. Since then the theory of shock acceleration in the interstellar medium has been developed fervently, and for the modern theory of this process readers are referred to the talk by Damiano Caprioli at this conference.

We notice, nevertheless, that the classical stochastic Fermi acceleration may also be effective in an astrophysical plasma. Stochastic acceleration may be effective near shocks of supernovae where magnetic turbulence  is generated by the Rayleigh-Taylor and Kelvin-Helmholtz instabilities, see e.g. \cite{yang}. Another example is the discovery of freshly accelerated CRs in the Cygnus Superbubble \cite{acker}. It was assumed, that CRs are accelerated there by the collective action of shocks in this area. The theory of stochastic (multi-shock) acceleration by a supersonic turbulence  was developed in \cite{bykov1,bykov}. 
    
\section{Theory of In-Situ Acceleration from a Background Plasma. The Number of Accelerated Particles}

One of the important questions, which models of acceleration should solve, is: how many high energy particles can be produced by these processes?  There
are no other  sources of particles for acceleration except those from a background plasma or high energy particles pre-accelerated by other sorces. We start from acceleration from a background plasma. 

 The spectrum of background plasma, $f_M$, is formed by Coulomb collisions and is described by the equation  (see \cite{lifsh})
\begin{equation}
\frac{1}{p^2}\frac{d}{dp}p^2\left[\left(\frac{dp}{dt}\right)_Cf_M+D_C\frac{df_M}{dp}\right]=0
\end{equation}
where  $p$ is the particle momentum, $(dp/dt)_C$ is the rate of Coulomb losses, 
$D_C=(dp/dt)_CmkT/p$ is the coefficient of momentum diffusion due to Coulomb collisions, $m$ is the mass of the accelerated particles, $k$ is the Boltzmann constant and $T$ is the plasma temperature. 

The solution of this equation is the
equilibrium Maxwellian distribution: 
\begin{equation}
f_M(p)=\sqrt{\frac{2}{\pi}}n_0\exp\left(-\frac{E}{kT}\right)
\label{fm}
\end{equation}
where $n_0$ is the density of background plasma and $E$ is the particle energy.

If background particles are under the influence of any acceleration, $(dE/dt)_{ac}=\alpha_0E$, (see Eqs. (2) and (3)) then the rate of energy 
variations is
\begin{equation}
\frac{dE}{dt}=\left(\frac{dE}{dt}\right)_{ac}-\left(\frac{dE}{dt}\right)_{C}
\label{en_var}
\end{equation}
where $({dE}/{dt})_{C}$ is the rate of ionization losses  in a gas with the density $n$, which  can be presented as (see e.g.  \cite{Ginz}) 
\begin{equation}
\left(\frac{dE}{dt}\right)_{C}=\nu_0E\left(\frac{E}{kT}\right)^{-3/2}
\end{equation}
where  
\begin{equation}
\nu_0=\frac{4\pi n_0e^4m_p^{1/2}\Lambda}{(kT)^{3/2}m_e}
\end{equation}
is the frequency of Coulomb collisions of thermal particles, $\Lambda$ is the Coulomb logarithm, $m_e$ and $m_p$ are the rest masses of electrons and protons.

Then Eq. (\ref{en_var}) has the form
\begin{equation}
\frac{dE}{dt}=\alpha_0E-
\nu_0E\left(\frac{E}{kT}\right)^{-3/2}\label{en_var1}
\end{equation}

From Eq.(\ref{en_var1}) we ca derive the threshold energy, $\varepsilon_{thr}$ 
\begin{equation}
\varepsilon_{thr}(n_0,\alpha_0) \simeq kT\left(\frac{\nu_0}{\alpha_0}\right)^{2/3}
\label{eq5}
\end{equation}
which defines the energy range of accelerated particles with $E>\varepsilon_{thr}$.

The stochastic acceleration forms a power-law spectrum of  nonthermal particles
\begin{equation}
f_{nth}=Kp^{-\gamma}
\label{fnth}
\end{equation}

The simplest way to estimate the number of accelerated particles (constant $K$) is just to match thermal (Eq. (\ref{fm})) and nonthermal (Eq. (\ref{fnth}))  components of the total spectrum at the energy $E=\varepsilon_{thr}$ that gives
\begin{equation}
K=\sqrt{\frac{2}{\pi}}n_0\exp\left(-\frac{\varepsilon_{thr}}{kT}\right)
\label{k0}
\end{equation}

However, as Gurevich noticed (see \cite{Gurev}),  the particle distribution
becomes non-equilibrium and time-varying in this case, and a proper estimate of the number of accelerated particles can be obtained in the framework of equation which includes the term of stochastic acceleration (described by another momentum diffusion with the coefficient  $D_F$) and the terms describing Coulomb collisions, which form the Maxwellian distribution of thermal particles. The total equation has the form
\begin{equation}
\frac{\partial f}{\partial t}-\frac{1}{p^2}\frac{d}{dp}p^2\left[\left(\frac{dp}{dt}\right)_Cf+(D_C+D_F)\frac{df}{dp}\right]=0
\label{k1}
\end{equation} 
that gives the solution for the distribution function of particles with e.g. $D_F=\alpha_0p^2$ as 
\begin{eqnarray}
&&N(\bar{p})=\sqrt{\frac{2}{\pi}}n_0\left[\exp\left(-\int\limits_0^{\bar{p}}\frac{udu}{1+(\alpha_0/\nu_o)\bar{p}^5}\right)\right.\nonumber\\
&&\left.-\exp\left(-\int\limits_0^\infty\frac{udu}{1+(\alpha_0/\nu_o)\bar{p}^5}\right)\right]
\label{k2}
\end{eqnarray}
where the dimensionless momentum $\bar{p}=p/\sqrt{2mkT}$.

\begin{figure}[h]
\begin{center}
\includegraphics[width=0.4\textwidth]{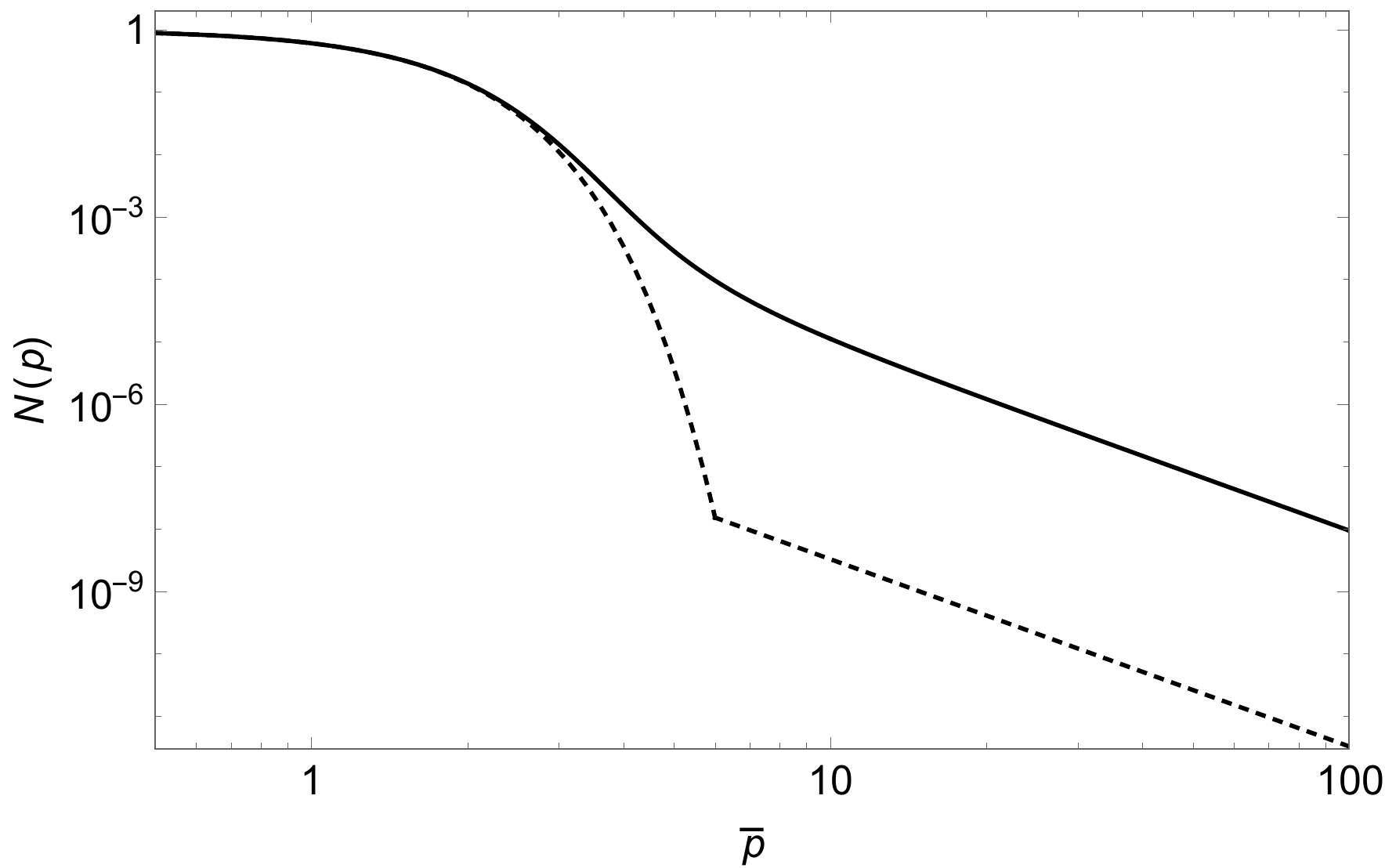}
\end{center}
\caption{Illustrative picture which shows the solution of correct kinetic equation (\ref{k2}) (solid line) and simple matching of thermal and nonthermal (power-law) components (dotted line). Here $\bar{p}$ is the dimensionless momentum $\bar{p}=p/\sqrt{2mkT}$. } \label{gur}
\end{figure}
The total spectrum of particles is shown in Fig. \ref{gur}. Two important conclusions  follow from this solution. The first one is that in the case of stochastic acceleration the Maxwellian and power-law components do not match with each other as assumed in Eq. (\ref{k0}). There is an extended region of a distorted Maxwellian distribution formed by Coulomb collisions, which tries to compensate the flux of particles running-away into the region of acceleration. Secondly, as
one can see from Fig. \ref{gur},  the simple estimate (\ref{k0}) underestimates strongly the number of accelerated particles.  

Dogiel et al.\cite{dog1,dog2} applied this model of particle acceleration from a background plasma for interpretation of the hard X-ray excess in the spectrum of the Coma cluster and showed that for reasonable parameters, this stochastic   acceleration is able to produce enough high energy particles needed to explain the Coma X-ray excess. 

However, in \cite{Gurev} a linear equation of acceleration was analysed with a constant temperature of plasma, $T=const$, which did not take into account a back reaction of accelerated particles onto  the parameter of the thermal pool. This was done by Wolfe and Melia\cite{wolfe} and Petrosian and East\cite{petr}, who showed from numerical calculations of a system of non-linear equations that this mechanism of in-situ acceleration did not work at all because the energy supplied by the acceleration  was immediately absorbed by the thermal pool, and the resulting effect of acceleration was a plasma overheating instead of a power-law spectra of non-thermal particles.

Later, Chernyshov et al.\cite{chern}  analysed a nonlinear system of equations describing acceleration and temperature variations. They showed that the resulting effect of in-situ acceleration from background plasma depended strongly on its parameters. If the momentum diffusion coefficient has a cut-off at low energies, e.g. in the form $D(p)=\alpha p^\zeta \theta(p-p_0)$ the situation depends drastically on the relation between the cut-off momentum, $p_0$ and the injection momentum, $p_{inj}=\sqrt{2m\varepsilon_{thr}}$. If $p_0<p_{inj}$ the effect of acceleration is similar to \cite{wolfe,petr}, i.e. the plasma is overheated. A surpisingly different result was obtained for the case of  $p_0>p_{inj}$. In this case acceleration subtracts from the thermal pool only high energy particles of the Maxwellian distribution. As a result the plasma cools down (analogue to Maxwell demon), and the power-law "tail" is formed by the acceleration. As a restriction of this model we should mention that the larger is the value  of $p_{inj}$, the smaller is the number of accelerated particles. On the other hand, for high enough $p_{inj}$ the Maxwellian spectrum matches directly with the power-law tail at $p_0$ instead of $p_{inj}$ as assumed in Eq. (\ref{k0}).

The question is what could be the reason for the acceleration cut-off at relatively low momenta. This could be due to absorption of MHD waves by CRs of relatively low energy \cite{ptus}. In the stationary case the equation for
spectrum of MHD-waves, $W(k,t)$  can be written as
\cite{norm}
\begin{equation}
\frac{d \Pi(W,k,t)}{dk}= -2\Gamma_{cr}W+\Phi\delta(k-k_0),
\label{Wk}
\end{equation}
where $k$ is the wave-number, $\Pi(W,k,t)$ decribes the non-linear cascade of waves, $\Phi$ is energy from  external sources at $k=k_0$, and
$\Gamma_{cr}$ is the decrement of absorption by CRs, see
\cite{ber90}
\begin{equation}
\Gamma_{cr}(k)=\frac{\pi Z^2
e^2V_A^2}{2kc^2}\int\limits_{p_{res}(k)}^\infty\frac{dp}{p}F(p)\,,
\end{equation}
where $p_{res}(k)=ZeH/ck$, $F(p)$
is the CR spectrum and $H$ is the magnetic field strength. 

The derived coefficient of the momentum diffusion is (see \cite{cheng})
\begin{equation}
D(p)= \alpha_0(p)J_2(\xi)
\end{equation}
where $\alpha_0(p)$ is a power-law function and $J_2$ is the Bessel function, where $\xi$ is a complicated function of $p$ (see for details of calculations Appendix in \cite{cheng}). The diffusion coefficient has a cut-off $D(p)=0$ at $J_2=0$, that corresponds to $\xi=5.14$ or  $p_0=0.2mc$ for  parameters of the Galactic halo .

\section{Stochastic Acceleration in the Galactic Halo. Models of the Fermi Bubbles}

Recent Fermi-LAT observations found new sources of CRs in the Galaxy whose origin is enigmatic. First of all, we mention   mysterious giant gamma-ray features in the central part of the Galaxy (Fermi Bubbles) elongated perpendicular to the Galactic plane\cite{dobl,su}. Several models were suggested to explain the origin of the bubbles which include  phenomenological assumptions about the processes of  particle acceleration there. Thus, Cheng et  al. \cite{cheng11} assumed that gamma-ray emission is generated by high energy electrons accelerated in the halo by  giant shocks resulting from tidal disruption of stars captured by the central black hole. Alternatively, Mertsch and Sarkar\cite{mertsch} assumed that this emission is produced by electrons in-situ accelerated by MHD-turbulence behind the shock. In this paper the authors tried to reproduce spectral characteristics of the emission from the Fermi bubbles but did not estimate whether this mechanism could provide enough electrons needed for the observed nonthermal fluxes from the bubbles. 

As we mentioned above there are no other evident sources of electrons for acceleration except those from the background plasma or those injected by supernova remnants. In the first case electrons are injected with energies close to their background temperature  in the halo (i.e. about keV). In the second case, only electrons with energies $E\leq 1$ GeV can reach the altitudes of the Fermi bubbles. In both case  further re-acceleration up to energies about $10^{12}$ eV is needed to generate gamma-rays from the FBs.

 Analysis of this acceleration for both situations was provided in \cite{cheng,cheng15}. In the first case, the kinetic equation for the distribution function is similar to Eq. (\ref{k1}) but it includes also synchrotron and inverse Compton energy losses for electrons. The number of accelerated electrons depends strongly on the value of cut-off momentum $p_0$. It cannot be too small because of the effect of plasma overheating, and too large because the number of accelerated particle is smaller than needed for the observed gamma-ray flux from the bubbles. In \cite{cheng} it was shown that  stochastic acceleration from a background is able to explain the observed emission from the Fermi bubbles but for an exceptionally narrow range of the acceleration parameters, which makes this model doubtful. 

The analysis of  the second case \cite{cheng15} showed that it is more effective for production of electrons than the acceleration from a background plasma, because in the case of SNR electron re-acceleration their energy should be increased by three orders of magnitude only while for acceleration from background pool electrons are accelerated from  their temperatures (about several keV). The kinetic equation in this case has a more complicated form than (\ref{k1}) because it includes also terms of particle propagation,
\begin{eqnarray}
&&-\nabla \cdot\left[\kappa(r,z,p)\nabla f -u(r,z)f\right]+\nonumber\\
&&\frac{1}{p^2}\frac{\partial}{\partial p}p^2\left[
\left(\frac{dp}{dt}-\frac{\nabla\cdot {\bf u}}{3}p\right)f -
D(r,z,p)\frac{\partial f}{\partial p}\right] =\nonumber\\
&&Q(p,r)\delta(z) \,,\label{eq_nu}
\end{eqnarray} 
where $r$ is the galactocentric radius, $z$ is the altitude  above
the Galactic plane, $p=E/c$ is the momentum of electrons, $u$ is the
velocity of the Galactic wind, $\kappa$ and $D$ are the spatial
and momentum (stochastic acceleration) diffusion coefficients,
$c(dp/dt)=dE/dt$ describes the rate of electron energy losses, and $Q$
describes the spatial distribution of cosmic ray (CR)
 sources in the Galactic plane ($z=0$) and their
injection spectrum. 

The problem of the second model is that the spectrum of re-accelerated electrons is too steep to reproduce the microwave emission from the bubbles as measured by Planck \cite{ade}. Thus, both models of stochastic acceleration of electrons in the bubbles have problems, and in that sense the model of electron acceleration by shocks \cite{cheng11} seems to be more attractive. We do not discuss here the hadronic model of Fermi bubbles (see e.g. \cite{crock11,fuji,thoudam13,Yang2014}), whose problems were presented  in \cite{cheng15b}

On the other hand, the stochastic acceleration of protons in the Fermi bubbles may explain the origin of CRs with energies above the "knee" ($>10^{15}$ eV) as shown by Cheng et al. in \cite{cheng12}. We presented, however, these results in details  at the last San-Vito conference which were published in \cite{chern14}.

\section{Stochastic Acceleration in the Galaxy and the problems Of CR Gradient}

The main questions of the theory of CR origin are where CRs are generated and how they propagate through the Galaxy. Necessary information can be obtained from investigations of the diffuse Galactic gamma-ray emission. First investigations of this emission showed that the derived distribution of CRs in the Galactic disk was flatter than the radial distribution of their potential sources: SNRs and pulsars (see e.g. \cite{strong88,strong96}). The first attempts to interpret this difference were performed in terms of CR propagation in the Galactic halo. In \cite{dog88,bloe93,erl13} it was assumed that an effective mixture of CRs in the Galactic halo due to CR scattering {diffusion}  made their distribution in the Galaxy more or less uniform. However, numerical calculations showed that even in the most favorable
case of an extended halo the diffusion  is unable to remove the signature
of the observationally inferred SNR source distribution. The problem is even more aggravated for the sharper SNR distribution of Green \cite{green}.

Recent analysis of the Fermi-LAT gamma-ray data \cite{acero16,yang3} in general confirmed a flatter CR distribution in the outer part of the Galaxy, although showed a sharp drop of CR density near the GC. Interpretation of the last result is beyond the scope of our analysis. If confirmed, it may be due to specific processes in the GC. Interpretation of the flat CR distribution in the outer Galaxy can be obtained in two different ways. One of them was suggested by \cite{dog88,strong00} who assumed that this is an effect of "unseen" SNRs at large galactocentric radii. 

 The other way was suggested by Breitschwerdt et al. \cite{Breit} and Recchia et al. \cite{recc16b} who interpreted the observed CR distribution in the Galactic disk in terms of  convective transport (galactic wind). In  \cite{Breit} the authors concluded from analyses of a system of hydrodynamic and kinetic equations that the wind velocity is proportional to the CR pressure $P_{CR}$ which in turn is proportional to the density of CR sources $Q_{CR}$. From analytical and numerical calculations of the three-dimensional equations of CR propagation similar to Eq. (\ref{eq_nu}), where the wind velocity is a function of coordinates and the source density, ${\bf u}(r,z,Q)$, they showed that CRs escaped faster from regions of higher source density. Just this effect explains in the model flatter CR distribution. 

In \cite{recc16b} the authors analysed a  non-linear model of CR propagation, in which their transport is determined by a self-generated turbulence. In their model they investigated a system of equations for  CR propagation (a one-dimensional version of Eq. (\ref{eq_nu}) for propagation in the direction perpendicular to the Galactic plane) and  equations for MHD-wave excitation by the CR streaming instability and their damping. In this model of non-linear CR transport the gradient and the spectral shape of CRs at different galactocentric radii  were reproduced. 

 If we return to models of CR acceleration we notice that at some conditions  the injection energy of stochastic acceleration is a function of plasma temperature. This may also be true for shock wave acceleration because in some models of particle injection into  shocks the temperature of the ambient
ISM has relevance \cite{bul,ber96,kang}. It is known that there is a radial increase of  temperature in HII regions  from about 6600K at $R = 2$ kpc to about 10000K at $R = 14$ kpc \cite{qui}. Recently, Erlykin et al. \cite{erl16} examined how many particles can be acclerated from background plasma depending on its temperature. 

\begin{figure}
\begin{center}
\includegraphics[width=0.4\textwidth]{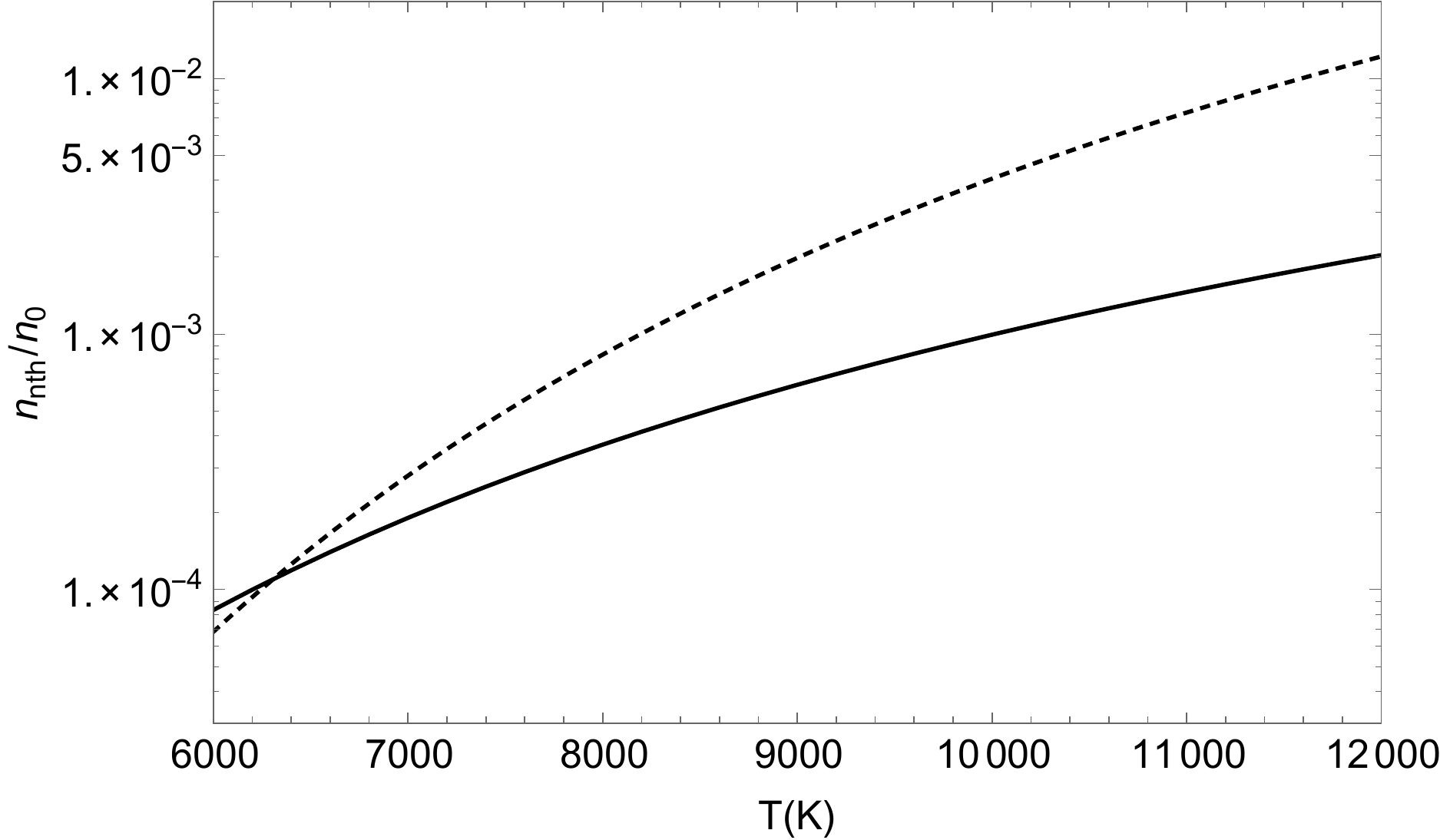}
\end{center}
\caption{Illustrative picture of the fraction of accelerated particles for stochastic (solid line) and shock wave (dotted line) acceleration as a function of temperature $T$ of background plasma. } \label{ratio}
\end{figure}

 For simple estimates of the fraction of  background particles accelerated by Fermi mechanism we take Eq. (\ref{k2}),  and for the  shock acceleration from background plasma we take the equation from  \cite{bul},
 which is 
\begin{equation}
\frac{\partial}{\partial x}\left(u(x)f-D\frac{\partial f}{\partial x}\right)=-\frac{1}{3}\frac{du}{dx}\frac{1}{p^2}\frac{\partial}{\partial p}(p^2f)
 \end{equation}
where the coordinate $x$ is perpendicular to the shock front, $D$ is the coefficient of
 particle spatial diffusion, $u$ is the particle fluid velocity and the velocity jump 
at the shock $(x=0)$ is 
$u_-/u_+=4$ for strong shocks with the Mach number $M>>1$. From this equation the number of accelerated particles can be estimated as
\begin{eqnarray}
&&\frac{n_{nth}}{n_0}\simeq \frac{p_T}{p_{thr}}\left(\frac{m_e}{m_p}\right)\delta^{1/3}\times\nonumber\\
&&\exp\left\{-\delta^{1/2}\left(1+\frac{1}{2}\ln\left[\left(\frac{m_i}{m_e}\right)^2\right]\frac{1}{3\delta}\right)\right\}
\label{ratio1}
\end{eqnarray}
where $m_e$ and $m_p$ are the electron and proton mass respectively, $m_i$ is the mass 
of the accelerated particles,  
\begin{eqnarray}
&&p_T=\sqrt{2kTm_i}\nonumber\\
&&p_{thr}=p_T\sqrt{\frac{m_p}{m_e}}\delta^{1/3}\nonumber\\
&&\delta=\frac{D\bar{\nu}}{u^2}\nonumber\\
&&\bar{\nu}=\frac{2\pi n_0e^4m_p^2}{m_e\bar{p}^3}\Lambda\nonumber\\
&&\bar{p}=m_p\sqrt{\frac{2kT}{m_e}}\nonumber
\end{eqnarray} 
As one can see in both cases the injection efficiency increases with the temperature $T$. In Fig. \ref{ratio} we presented the fraction of accelerated particles as it follows from Eqs. (\ref{k2}) and (\ref{ratio1}).

As it follows from the radio data, the density of SNRs drops by 30 times from the radius $R=2$ kpc to $R=14$ kpc \cite{green}, while the density of CRs drops by 2 times only for these radii. However, as we see in Fig. \ref{ratio}, the model efficiency of particle acceleration rises in 20-100 times for these distances from the GC because of the temperature variations. Thus, the efficiency of acceleration may compensate partially the drop of supernova density and with this effect the observed CR density can be reproduced in the "temperature" model. Erlykin et al. \cite{erl16} derived from the Maxwell-Boltzman distribution at different temperatures  the necessary value of $\varepsilon_{thr}$, which should be about 2.5 eV  for the observed CR densty variations.  This analysis   gives a reasonable coincidence of the "temperature" model with variations of CR density in the Galactic disk although  there are a number of uncertainties which require further investigations.

We notice, however, that this estimates presented in Fig. \ref{ratio}, are mainly illustrative and give qualitative impression about the injection processes. Thus,  shocks in the Galaxy are collisionless, and particle injection is determined by interactions with magnetic fluctuations (not by Coulomb collisions, see the talk of Damiano Caprioli).

\section{Conclusion}
We  give a short review of process stochastic acceleration in the Galaxy. The conclusions are itemized below:
\begin{itemize}
\item In the case of stochastic acceleration from a background plasma the distribution is nonequilibrium. Coulomb collisions try to compensate a flux of particles running-away from a background pool into the acceleration region. Therefore, a very broad transfer  region is formed by Coulomb collisions between thermal and non-thermal components of the total spectrum. In order to estimate the number of accelerated particles  one should analyse a kinetic equation which includes both the term of Coulomb collisions, which forms the Maxwellian distribution of thermal particles, and the term of stochastic acceleration, which forms a nonlinear power low spectrum of non-thermal particles.
\item If the stochastic acceleration interacts with all particles of the Maxwellian spectrum, then it does not form power-law "tails" of accelerated particles. The energy supplied by acceleration is absorbed immediately by the thermal pool through ionization losses of accelerated particles. The resulting effect of stochastic acceleration is plasma overheating.
\item If the stochastic acceleration interacts  with a high energy fraction of the Maxwellian distribution only, then power-law spectra of particles are generated, and the number of accelerated particles depends strongly on the position of acceleration the low momentum cutoff of the stochastic acceleration process.
\item We analysed whether the stochastic acceleration in the Galactic halo can produce there enough number of high energy electrons needed to explain the gamma-ray and microwave emission from the enigmatic Fermi bubbles in the Galactic central region. We analysed two cases of electron acceleration: a) in-situ acceleration of electrons from background plasma, and b) re-acceleration in the halo of electrons generated by SNRs in the disk, which reach the altitudes of the Fermi bubble edges. We showed that there are problems in both cases, but the needed number of electrons can be provided under  specific conditions.
\item We discussed whether the CR gradient in the Galactic disk can be explained in terms of the model of stochastic acceleration. We showed that if the number of accelerated particles is a function of the temperature of background plasma then this effect may explain the observed radial variation of the CR density in the Galactic disk.
\end{itemize}

\section*{Acknowledegments}
V.A.D. and D.O.C. acknowledge a partial support from the MOST-RFBR grant 15-52-52004 and the RFBR grant 15-02-02358.   K.S.C. is supported by the  GRF Grants of the Government of the Hong Kong SAR under HKU 17310916. C.M.K. is supported in part by the Taiwan Ministry of Science and Technology Grants MOST 104-2923-M-008-001-MY3 and MOST 105-2112-M-008-011-MY3. A.D.E. and A.W.W. are grateful to the Kohn Foundation for the financial support (Grant RF 040081).




\nocite{*}
\bibliographystyle{elsarticle-num}
\bibliography{jos}



\end{document}